\documentclass{article}
\usepackage[utf8]{inputenc}
\usepackage[final, nonatbib]{neurips_2024}

\title{Transforming Simulation to Data Without Pairing}
\usepackage[backend=bibtex, sorting=none]{biblatex}
\bibliography{bib}

\usepackage[T1]{fontenc}    
\usepackage{hyperref}       
\usepackage{url}            
\usepackage{booktabs}       
\usepackage{amsfonts}       
\usepackage{nicefrac}       
\usepackage{microtype}      

\usepackage{rotating}
\usepackage{array}
\usepackage{amsmath}
\usepackage[normalem]{ulem}
\usepackage{slashed}
\usepackage[pdftex,table]{xcolor}
\usepackage{units}
\usepackage{xfrac}
\usepackage{mathtools}
\usepackage{empheq}
\usepackage{amssymb}
\usepackage{url}
\usepackage{comment}
\usepackage{paralist}

\usepackage{float}

\def\beq{\begin{equation}}
\def\eeq{\end{equation}}
\newcommand{\bea}{\begin{eqnarray}\begin{aligned}}
\newcommand{\eea}{\end{aligned}\end{eqnarray}}
\def\bitem{\begin{itemize}}
\def\eitem{\end{itemize}}

\definecolor{darkpurple}{rgb}{0.5, 0.2, 0.8}
\definecolor{darkblue}{rgb}{0.0, 0.0, 0.8}
\definecolor{darkgreen}{rgb}{0.0, 0.4, 0.0}
\definecolor{darkred}{rgb}{0.5, 0.0, 0.0}
\usepackage{hyperref}
\hypersetup{
    linktocpage,
     colorlinks,
     citecolor=darkgreen,
     linkcolor= darkgreen,
     urlcolor=darkgreen
}

\author{%
  Eli Gendreau-Distler$^1$ \\
  \texttt{egendreaudistler@berkeley.edu} \\
  \And  
  Luc Le Pottier$^{1}$ \\
  \texttt{luclepot@berkeley.edu} \\
  \And
  Haichen Wang$^{1,2}$ \\
  \texttt{haichenwang@lbl.gov} \\
  \\
  $^1$ Physics Department, University of California, Berkeley\\
  Berkeley, CA 94720 USA \\
  $^2$ Physics Division, Lawrence Berkeley National Laboratory\\
  Berkeley, CA 94720 USA \\
}

\begin{document}

\maketitle

\begin{abstract}
We explore a generative machine learning-based approach for estimating multi-dimensional probability density functions (PDFs) in a target sample using a statistically independent but related control sample—a common challenge in particle physics data analysis. The generative model must accurately reproduce individual observable distributions while preserving the correlations between them, based on the input multidimensional distribution from the control sample. Here we present a conditional normalizing flow model ($\mathcal{CNF}$) based on a chain of bijectors which learns to transform unpaired simulation events to data events. We assess the performance of the $\mathcal{CNF}$ model in the context of LHC Higgs to diphoton analysis, where we use the $\mathcal{CNF}$ model to convert a Monte Carlo diphoton sample to one that models data. We show that the $\mathcal{CNF}$ model can accurately model complex data distributions and correlations. We also leverage the recently popularized Modified Differential Multiplier Method  (MDMM) to improve the convergence of our model and assign physical meaning to usually arbitrary loss-function parameters.
\end{abstract}

\section{Introduction}\label{sec:intro}
In particle physics data analysis, a ubiquitous challenge involves determining the correlated distributions of multiple observables within a target sample for a specific physics process. This problem centers on estimating a multi-dimensional probability density function (PDF). Traditionally, this estimation relies on extrapolating or interpolating from another multidimensional PDF derived from a statistically independent but related control sample. This extrapolation generally employs explicit knowledge of the physics underlying the relationship between the two samples. Generative machine learning enables a novel method to learn these relationships between multidimensional PDFs across different samples. Generative models typically convert a base distribution into a target distribution, mirroring the process of extrapolating multi-dimensional PDFs from a control sample to the target sample. Compared to traditional methods, generative machine learning approaches are expected to uncover more intricate relationships, particularly the correlations between observables. In this paper, we explore this methodology using a conditional normalizing flow model, looking in particular at the case where we want to transform unpaired Monte Carlo (MC) simulation samples into data samples at the level of high-level analysis features.


\section{Methods}\label{sec:methods}
\subsection{Datasets}\label{sec:data}
The training datasets for this study include both a Monte Carlo (MC) simulation and ATLAS collider data from CERN OpenData. The ``MC'' sample is an $H \rightarrow \gamma\gamma$ production sample at $\sqrt{s} = 13$ TeV, using Madgraph@NLO v2.3.7 \cite{Alwall:2014hca} for event generations at next-to-leading order accuracy and Pythia 8.234 \cite{Sjostrand:2014zea} with the CTEQ6L1 parton distribution function set \cite{pumplin_new_2002}. The ``data'' sample is an ATLAS OpenData sample with 10fb$^{-1}$ of pp collision data at $\sqrt{s} = 13$~TeV \cite{atlas_collaboration_atlas_2020}. Both datasets have identical feature selection requiring exactly two photons, with combined invariant mass $60 \leq m_{\gamma\gamma} \leq 300$ GeV, leading photon transverse momentum $35 \leq p_{T_1} \leq 250$ GeV, subleading photon transverse momentum $25 \leq p_{T_2} \leq 250$ GeV, angular coordinates $|\eta| \leq 1.37$ or $1.52 \leq |\eta| \leq 2.37$, and $|\phi| \leq \pi$. The total post-selection size for both datasets was about 6.5 million events. The final training events consisted of $p_T$, $\eta$, and $\phi$ for each photon, totaling 6 features per event. Feature distributions for both data and MC samples are displayed in Figure~\ref{fig:features}, while correlations are shown in Table~\ref{tab:corr}.



\subsection{Conditional Normalizing Flows}\label{sec:cnf}
A normalizing flow transforms a simple base density distribution $\pi\left(\vec{z}\right)$ to a target density distribution $p\left(\vec{x}_\text{Ref}\right)$ using a learnable, invertible mapping $f_\phi$ and applying the change of variables formula~\cite{rezende_variational_2015}. 
Normalizing flows are then typically trained by minimizing the negative log-likelihood function $\mathcal{L}\left(\vec{w} | \vec{x}_\text{Ref}\right) = -\mathbb{E}_x\left[\log{\left(p_w\left(\vec{x}_\text{Ref}\right)\right)} \right]$ for model parameters $\vec{w}$. Conditional normalizing flows ($\mathcal{CNF}$) extend this framework and can perform density estimation dependent on a \textit{conditional vector} $\vec{x}_c$, allowing estimation of the conditional density distribution $p(\vec{x}_\text{Ref} | \vec{x}_c)$~\cite{winkler_learning_2023}. Our $\mathcal{CNF}$ implementation is based on Masked Autoregressive Flows (MAFs)~\cite{papamakarios_masked_2017} with a Gaussian base distributions. In this paper we use the 6 kinematic features described in section~\ref{sec:data} for both the conditional vector $\vec{x}_c$ and the target distribution $\vec{x}_\text{Ref}$. We also use simulated MC events for $\vec{x}_c$, and unpaired ATLAS OpenData events for $\vec{x}_\text{Ref}$.

The target distribution $p(\vec{x}_\text{Ref} | \vec{x}_c)$ is then learned such that $\vec{x}_c + p(\vec{x}_\text{Ref} | \vec{x}_c) \approx \vec{x}_\text{Ref}$, or equivalently $p(\vec{x}_\text{Ref} | \vec{x}_c) \approx \vec{x}_\text{Ref} - \vec{x}_c \equiv \vec\Delta_c$, where the ``delta'' values parameterize the difference between the conditional vector $\vec{x}_c$ and the reference distribution $\vec{x}_\text{Ref}$. Because there is not any relationship between individual events in the MC and data, the adjusted MC, $\vec x_c + \vec\Delta_c$, will not match the data on an event-by-event basis. Across the entirety of both distributions, however, we can learn to predict $\vec\Delta_c$ such that the overall distribution of the transformed MC and the data are nearly identical. 


\vspace{-10pt}
\subsection{Loss Functions}
Given our unique requirements for distribution similarity, we propose and evaluate novel combinations of the following loss functions in addition to the typical $\mathcal{CNF}$ loss function described in section~\ref{sec:cnf}.

\begin{figure}[h]
\centering
\centerline{\includegraphics[width=\textwidth]{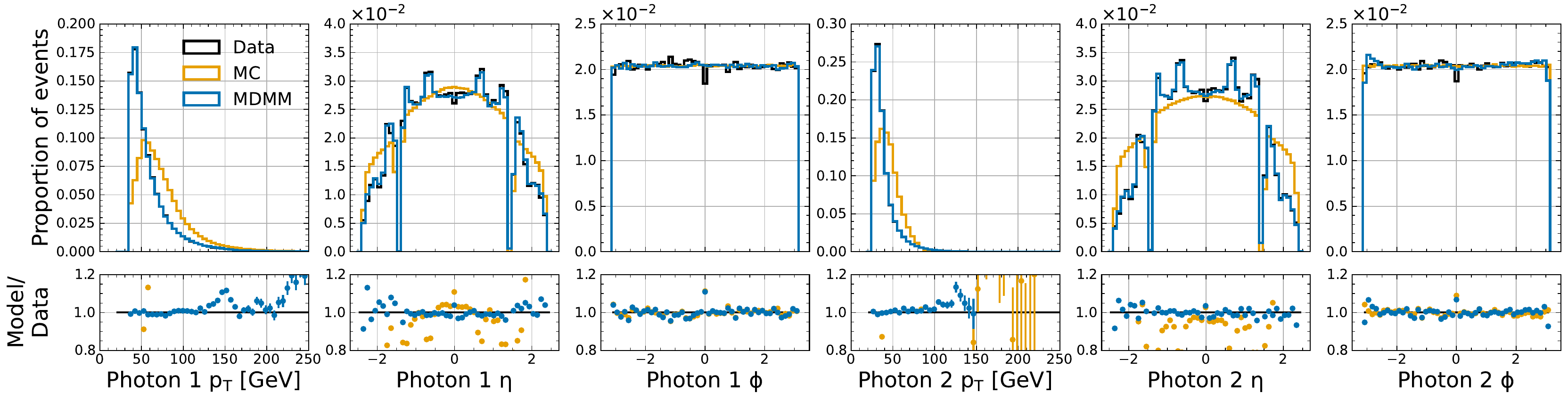}}
\caption{Leading and subleading photon $p_\mathrm{T}, \eta,$ and $\phi$ from MC, data, and generated samples using MDMM with MMD/KL divergence losses.}
\label{fig:features}
\end{figure}

\subsubsection{Kullback-Leibler Divergence (KL Divergence)}\label{sec:kl}

Kullback-Leibler (KL) divergence is a distance metric which operates on two probability distributions $P$ and $Q$, and measures the difference in these probability distributions \cite{kullback_information_1951}. It is defined as
\begin{equation}\label{eq:kl}
    D_{\text{KL}}\left(P \parallel Q \right) = \sum_{x \in \mathcal{X}} P(x) \log{\left(\frac{P(x)}{Q(x)}\right)}
\end{equation}

In the case that the KL divergence is evaluated across all features, minimizing it is an optimization problem equivalent to maximizing the likelihood, as is typically done to train a $\mathcal{CNF}$. But the KL divergence can also be evaluated across any subset of the training features, to different effects as a loss term. 

One relevant usage in HEP analyses might be to improve the modeling of very fine distribution details. In this paper, these details come in the $\eta$ distributions of photons from Higgs decays. These features, shown in Figure~\ref{fig:features}, typically contain complicated details caused by a variety of position-dependent ``detector effects." Such effects vary, but in ATLAS they may include the amount of detector material between the collision and detection points, varying detector efficiencies, or highly non-uniform radiation damage, among many others.

In order to construct a loss term focused on the one-dimensional projections of our features, we compute the KL divergence on the one-dimensional marginals, rather than on the full feature set. This means that we make a separate KL evaluation for each of our $N$ features, resulting in $N$ 1-dimensional KL evaluations rather than one $N$-dimensional KL evaluation.

To ensure we have enough statistics to make such a calculation viable, we take a batch size of 25600 events per training step. From this we calculate a Gaussian Kernel Density Estimate (KDE) on each of our input features. Gaussian KDEs provide robust density estimation for low-dimensional distributions and can be tuned with a bandwidth parameter, which is set for each feature individually prior to training~\cite{rosenblatt_remarks_1956, parzen_estimation_1962}. KDEs also have the added benefit of producing a continuous function, as well as a bandwidth parameter for tuning how finely the features are to be estimated. KDEs for the reference distribution are re-generated at each epoch, while KDEs for the entire target dataset are precomputed at the start of training. Once the KDEs are calculated we then sample the PDFs for each feature, calculate the KL divergence between the modified reference distribution and the target distribution, and average over all training features. Further tuning of the bandwidths of the KDEs and the weights for each feature can also be applied, but is not done in this work.

\subsubsection{Maximum Mean Discrepancy (MMD)}\label{sec:mmd}

While KL divergence provides an excellent metric for each feature distribution shape, it does not encourage the model to learn correlations between feature variables. We introduce here an alternate loss function which does learn correlations; the Maximum Mean Discrepancy (MMD)~\cite{gretton_kernel_2006, gretton_kernel_2012}. MMD is calculated by evaluating a kernel function for all pairs of output-output, output-target, and target-target samples in the MC dataset $\vec{x}_\text{MC}$ and in the Data $\vec{x}_\text{Data}$. Using the Gaussian kernel function $k_\sigma(x_1,x_2)=\exp{\left(-\frac{1}{\sigma}\parallel x_1-x_2 \parallel^{2}\right)}$ with bandwidth $\sigma$, we define MMD to be 
\begin{equation}
    \text{MMD}_\sigma = 
    \frac{1}{n^2}\sum_{i=1}^n \sum_{j = 1}^n k_\sigma\left({\vec{x}_\text{MC}}^{(i)}, {\vec{x}_\text{MC}}^{(j)}\right) +
    \frac{1}{n^2}\sum_{i=1}^n \sum_{j = 1}^n k_\sigma\left({\vec{x}_\text{Data}}^{(i)}, {\vec{x}_\text{Data}}^{(j)}\right) -
    \frac{1}{n^2}\sum_{i=1}^n \sum_{j = 1}^n k_\sigma\left({\vec{x}_\text{MC}}^{(i)}, {\vec{x}_\text{Data}}^{(j)}\right)
\end{equation}

We use the same bandwidth for all features of $\sigma = 0.1$, calculated across batches of $n = 1000$ events, and averaged over all features. Similar to the KDE, this bandwidth may also be tuned.

\begin{figure}[H]
\centering
\begin{minipage}{0.32\textwidth}
    \includegraphics[height=.93\textwidth]{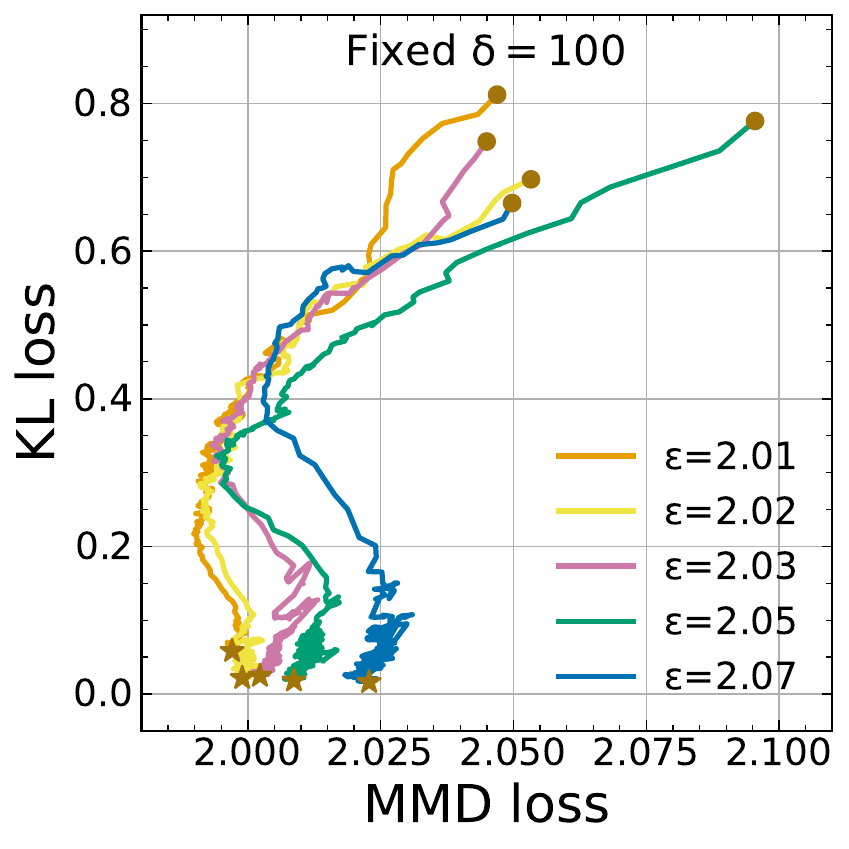}
\end{minipage}
\hfill
\begin{minipage}{0.32\textwidth}
    \includegraphics[height=.93\textwidth]{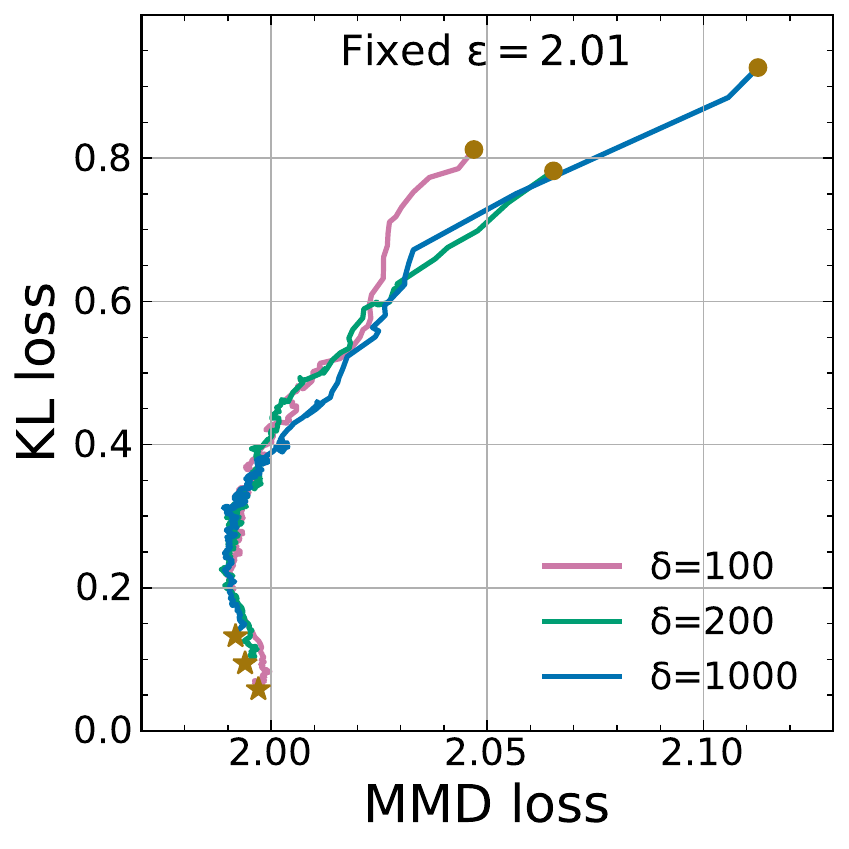}
\end{minipage}
\hfill
\begin{minipage}{0.34\textwidth}
    \includegraphics[height=.87\textwidth]{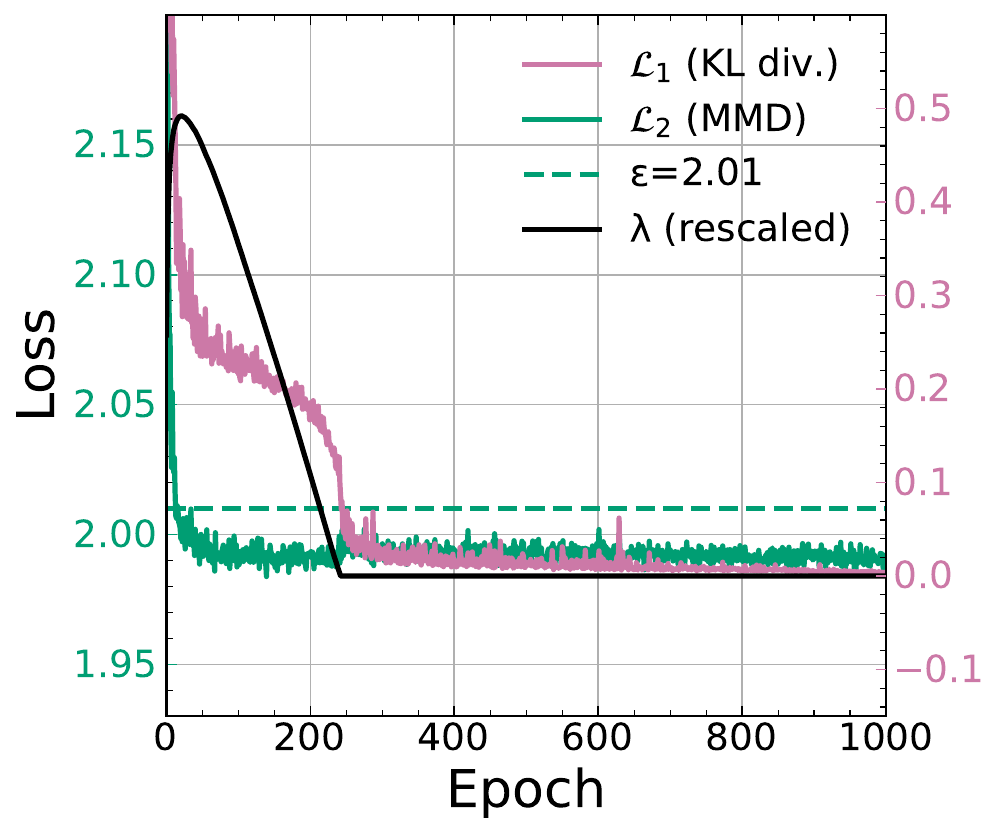}
\end{minipage}
\caption{Left/Center: Trajectories of MDMM models over 500 epoch trainings for either fixed $\delta$ (left) or $\epsilon$ (right). Right: Evolution of primary and secondary losses \(\mathcal{L}_1\) and \(\mathcal{L}_2\) and learned hyperparameter \(\lambda\) over 1000 epoch training of MDMM model with constraint \(\mathcal{L}_2< \epsilon=2.01\).}
\label{fig:mdmm}
\end{figure}
\subsection{Modified Differential Multiplier Method}\label{sec:mdmm}

To arrive at an optimal balance of influences for multiple loss functions, we use the Modified Differential Method of Multipliers (MDMM) algorithm introduced in Ref.~\cite{platt_constrained_1987}. MDMM identifies primary and secondary loss functions $\mathcal{L}_1(\vec w)$ and $ \mathcal{L}_2(\vec w)$, each depending on model parameters $\vec w$, and reformulates the loss function as a constrained optimization problem:
\begin{equation}
    \mathcal{L}(\vec w, \lambda) = \mathcal{L}_1(\vec w) - \lambda(\epsilon - \mathcal{L}_2(\vec w)) + \delta(\epsilon - \mathcal{L}_2(\vec w))^2
\end{equation} 
where $\lambda$ is a learned parameter dynamically updated during training, $\epsilon$ is the target loss value for $\mathcal{L}_2(\vec w)$, and $\delta$ is a damping parameter to tune the rate of convergence. In this work we take the KL loss to be the primary loss function $\mathcal{L}_1(\vec w)$ and the MMD loss to be the secondary loss function $\mathcal{L}_2(\vec w)$. We take the target MMD loss ($\epsilon = 2.01$) to be slightly larger than the MMD between different batches of the target dataset ($1.98$). Figure~\ref{fig:mdmm} shows the results of scanning over values of $\delta$ and over values of $\epsilon$ close to the data-vs-data MMD value, as well as a plot of the convergence of the weighting parameter $\lambda$ over the training lifetime for a single model. Using MDMM, we are able to automatically learn $\lambda$ so as to optimize the trade-off between the primary and secondary loss functions. 


\section{Results}\label{sec:results}

We present here the results for a normalizing flows model trained with four different loss functions: the basic $\mathcal{CNF}$ Log Loss (section~\ref{sec:cnf}), KL divergence only (sec~\ref{sec:kl}), MMD only (sec~\ref{sec:mmd}), and finally a combination of KL divergence and MMD, balanced using the MDMM (sec~\ref{sec:mdmm}). 

The training/validation split was $90\%/10\%$, and the model consisted of 10 bijector MAF blocks, each of which contains two dense layers with latent size 128 and ReLU activation functions. The batch size was 25600 for all trainings except those with MMD, in which case a batch size of 1000 was used to avoid memory overflow issues. We used the RMSProp optimizer~\cite{adam}, with a square root learning rate decay decreasing from 10$^{-3}$ to 10$^{-5}$ over the course of the 1000-epoch training time. Further details can be found in the code accompanying this paper\footnote{Full code used for this paper can be found at \href{https://gitlab.cern.ch/egendrea/nf_for_modeling}{https://gitlab.cern.ch/egendrea/nf\_for\_modeling}}.

\begin{figure}[H]
\centering
\centerline{\includegraphics[height=.38\textwidth]{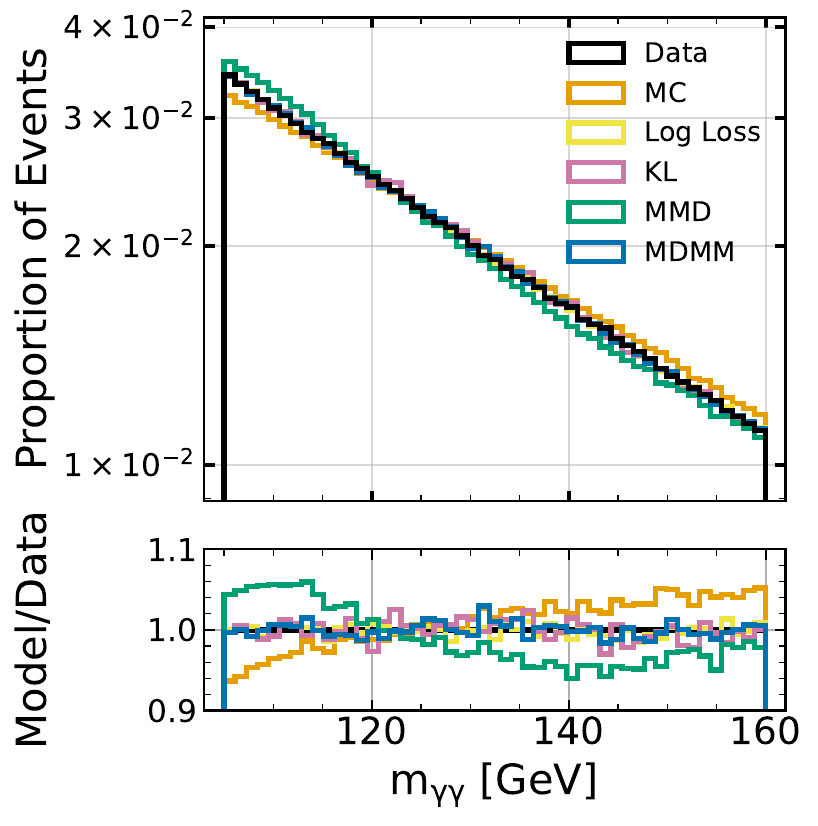}\includegraphics[height=.38\textwidth]{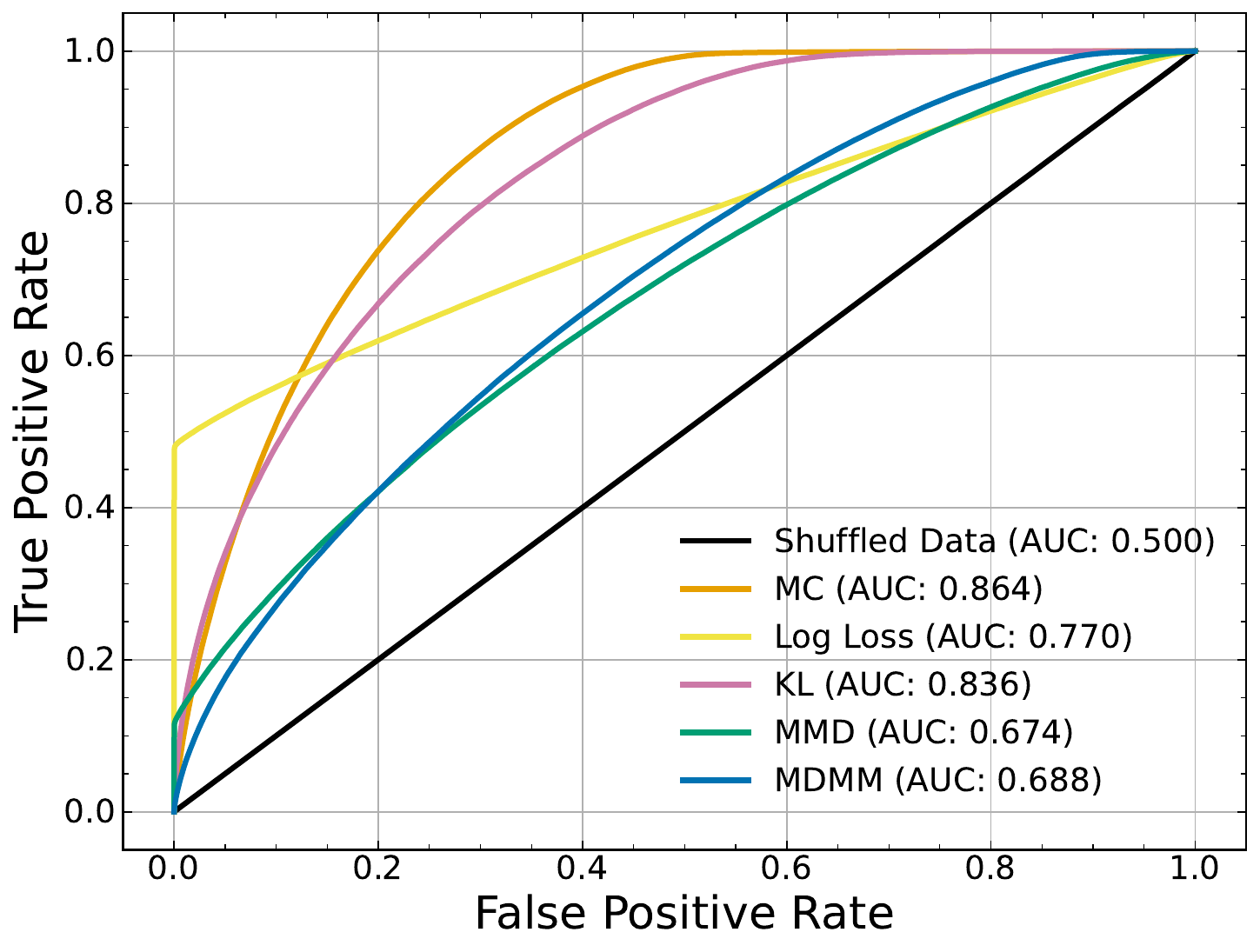}}
\caption{Left: $m_{\gamma\gamma}$ distributions for OpenData, MC, and generated samples. Right: ROC curves for a classifier trained to distinguish data from shuffled data, MC, and generated samples. Worse performance here indicates better generative model performance.}
\label{fig:perf}
\end{figure}

As a first test, we show the $m_{\gamma\gamma}$ distributions for MC, data, and all generative models in Figure~\ref{fig:perf}. Accurate modeling of this distribution is crucial for the \( H \rightarrow \gamma\gamma \) analysis, and it is imperative that any generative model be able to faithfully replicate this distribution. It is clear that the MDMM model in particular is able to reproduce the distribution quite precisely, especially in the signal region $100 < m_{\gamma\gamma} < 160$ GeV.

Feature distributions for MC, data, and the MDMM generative model are shown in Figure~\ref{fig:features}, and showcase how well the MDMM model is able to capture fine distribution structure such as detector effects in the photon $\eta$ distributions. This is in contrast to any of the other loss functions, which only reproduce the simulation in the $\eta$ feature, at best (these are excluded for clarity of presentation). Table~\ref{tab:corr} displays a table of Pearson correlation coefficients, from which we can clearly see that the models trained with MMD and Log Loss match much better with the target ATLAS data than any of the other models.

\begin{table}[H]\label{tab:corr}
\caption{Pearson correlation coefficients for data, MC, and generated samples. Columns 2-6 list all features which are correlated in data, while the mean value of all others is listed in the final column.}
\begin{center}
\scalebox{0.75}{
\begin{tabular}{|c|c|c|c|c|c|c|c|}
    \hline
    & & & & & & {Mean Abs. Value}\\
    Dataset & $p_{T_1}$, $p_{T_2}$ &	
    $\eta_1, \eta_2$ &	$\phi_1$, $\phi_2$ &
    $m_{\gamma\gamma}$, $p_{T_1}$ &	
    $m_{\gamma\gamma}$, $p_{T_2}$ &	
    {of All Others} \\
    \hline
    {\textit{ATLAS Data}} & \textit{0.539}&	\textit{0.192}&	\textit{-0.425}&	\textit{0.566}&	\textit{0.589} & \(\textit{7.06}\cdot \textit{10}^{\textit{-4}}\)\\
    {MC} & 0.103&	0.459&	-0.394&	0.356&	0.417& \(3.17\cdot10^{-4}\)\\
    {Log Loss} & 0.542&	0.189&	-0.422&	0.559&	0.583& \(7.43\cdot 10^{-4}\)\\
    {KL} & 0.542&	0.197&	-0.421&	0.591&	0.601& \(1.74\cdot 10^{-1}\)\\
    {MMD} & 0.535&	0.189&	-0.425&	0.572&	0.580& \(2.62\cdot 10^{-3}\)\\
    {MDMM (MMD/KL)} & 0.550&	0.203&	-0.422&	0.558&	0.584& \(4.20\cdot 10^{-3}\)\\
    \hline
\end{tabular}
}
\end{center}
\end{table}

As a final comparison between the four $\mathcal{CNF}$ models, we trained a discriminating neural network to distinguish data from generated samples. The discriminator consisted of 2 layers of 32 nodes each, using the 6 kinematic features as input, with a batch size of 1000 events and Binary Crossentropy (BCE) loss. The Receiver Operating Characteristic plots for this discriminator when trained on various $\mathcal{CNF}$ models are plotted in the right side of Figure~\ref{fig:perf}. We can see that the MDMM model outperforms the MC, but has comparable performance to a model trained with MMD only.

\section{Conclusions}\label{sec:conclusions}

We have presented a variety of methods for transforming MC simulation event samples to real data samples using $\mathcal{CNF}$ trained on analysis-level features for each event. Using the specific analysis example of $H\rightarrow\gamma\gamma$ decays, we present a variety of metrics such as visual distribution similarity, correlations, and discriminant AUC scores. We show how different loss terms may be leveraged for different analysis purposes -- for instance, while MMD alone is equivalent to MMD + KL divergence combined using the MDMM in terms of AUC performance, visually the $\eta$ distributions have much finer modeling in the latter case.

Using this method, we obtain a generalized transformation function to convert simulated samples to data-like samples for specific analysis cases. Since this method still requires input simulation samples, it  does not provide a directly ``generative'' model, but instead yields a transformation applying to the particular analysis it was trained on. One interesting use case for this type of model is for analyses targeting rare processes, which can require the generation of orders of magnitude more MC statistics than will actually pass the analysis cuts. In this case other NN-based re-weighting methods for improving MC predictions may be challenging to train. Using our proposed method, one could train a $\mathcal{CNF}$ on higher-statistics control regions (CRs) or pre-cut MC, before evaluating on the signal region MC data, providing tunable re-weighting without needing high statistic samples. 

Future studies could explore more analysis cases, different conditional distribution modeling methods (GAN, VAE, etc.) instead of $\mathcal{CNF}$, testing of new loss functions, and testing the ability of this method to generalize to different MC/data samples with similar target processes. 



\printbibliography

\end{document}